\documentclass[preprintnumbers,a4paper,twocolumn,10pt,showpacs,prl]{revtex4}%
\usepackage{graphicx}
\usepackage{dcolumn}
\usepackage{bm}
\usepackage{latexsym,amsthm,amssymb}
\usepackage{booktabs,threeparttable}
\usepackage{makeidx}
\usepackage{amsmath}
\usepackage{amsfonts}
\usepackage{amssymb}%
\providecommand{\U}[1]{\protect\rule{.1in}{.1in}}
\begin{document}
\title{\textbf{Long range orders in multiferroics}}
\author{C. D. Hu}
\affiliation{Department of Physics and Center for Theoretical Sciences}
\affiliation{National Taiwan University}
\affiliation{Taipei, Taiwan}
\affiliation{Republic of China}
\date{\today }

\begin{abstract}
We proposed that in multiferroics there exists a third long range order
besides the electric polarization and magnetic order. This long range order
reduces the symmetry of the spatial part of the wave functions of electrons.
Thus the cancellation in the "spin current" model can be avoided. As a result,
the expectation value of electric polarization will be larger by an order of
magnitude. We have derived a new form of electric polarization $\vec{P}%
\sim-\widehat{Q}\times(\widehat{s}_{j}\times\widehat{s}_{j+1})$\ where
$\vec{Q}$\ is the wave vector of this long range order.
\end{abstract}

\maketitle

Multiferroic materials \cite{1} have several coexisting long range orders.
Kimura et.\ al. \cite{3} found very interesting properties in TbMnO$_{3}$. The
ferroelectric transition temperature coincides with a magnetic transition
which was identified later to be the transition from sinusoidal magnetic order
to spiral order \cite{4}. It was also found \cite{5} that with sufficiently
strong magnetic field, the electric polarization switches direction. These are
clear experimental evidences that the electric polarization and magnetic
orders are closely related. Electromagnetic field is one of the fundamental
interactions and electricity and magnetism are unified. It should not have
been surprising that electric and magnetic properties are coupled. However, in
this case there seems to be no charge current nor time-dependent electric
field, nor varying magnetic flux in the system. So the fundamental question is
by what mechanism electricity and magnetism are coupled.
As the list of multiferroic material grows rapidly, their mechanisms and the
varieties of physical properties also multiply \cite{9,10}. Van der Brink and
Khomskii classified them into two types \cite{11}. Type-I multiferroics
usually have high ferroelectric temperature but weak coupling between electric
polarization and magnetic order. In type-II multiferroics, there are strong
electric polarization and magnetic order coupling.\ With a few exceptions such
as Ca$_{3}$Co$_{2-x}$Mn$_{x}$O$_{6}$ \cite{12} and the scenario proposed by
Sergienko et. al., \cite{13}, these compounds are originated from spin-orbit
interaction and possess spiral spin configuration. This is the focus of this work. Katsura et. al. (KNB) \cite{14} proposed the spin current model of
multiferroics in which the coupling between electric polarization and magnetic
order is due to spin-orbit interaction. The mechanism can be interpreted as
the coupling between electric field and spin current, or the Aharonov-Casher
(AC) effect \cite{15} and it gives polarization
\begin{equation}
\vec{P}\simeq-eI\left(  \frac{V}{\Delta}\right)  ^{3}\widehat{e}_{12}%
\times(\widehat{e}_{1}\times\widehat{e}_{2}) \label{eq1}%
\end{equation}
where $V$\ is the hybridization energy, $\Delta$ is the charge transfer
energy, $\widehat{e}_{1}$\ and\ $\widehat{e}_{2}$ are the unit vectors of two
adjacent spins and $\widehat{e}_{12}$ is the direction of bond.
$eI=e\left\vert {\int}\psi_{d}^{\ast}(\vec{r})\vec{r}\psi_{p}(\vec{r}%
)d^{3}\vec{r}\right\vert $, is the expectation value of electric polarization
of the hybridized state of d-orbitals of transition element atoms and
p-orbitals of oxygen atoms. The factor $\widehat{e}_{1}\times\widehat{e}_{2}$
fits nicely with spiral spin configuration. It predicts that the electric
polarization is perpendicular to the screw direction. Furthermore, it was able
to explain the results of Kimura et. al. \cite{5}: applied magnetic field can
flip the direction of spiral spins and hence, that of electric polarization.
The generalization to a bulk system with active e$_{g}$ orbitals was also
performed \cite{16}.

Though the spin current model has been successful in many aspects, it has the
quantitative problem. For example, assuming $V\approx0.5eV$\ and
$\Delta\approx2eV$, $I$ has to be of the order $0.1nm$\ in order that
$P\approx1000\mu C/m^{2},$\ the magnitude in TbMnO$_{3}.$ This raised doubt of
the relevance of the model \cite{17,18}. First-principle calculation by
Malashevich and Vanderbilt \cite{19} indicated that in TbMnO$_{3}$ the
electric polarization of purely electronic nature is an order of magnitude
smaller than experimental data. This problem has to be addressed before the
spin current model can be viewed as an accomplished\ model.\ This is also the
subject of this work.

The electric polarization in Eq.(\ref{eq1}) which has the factor $(V/\Delta)$
to the third power, is the remains of catastrophic cancellation which comes
from the symmetry of the system. In KNB's words: "the dominant term comes from
the difference of the normalization factors between two perturbed states." The
electric polarization exists only in the systems without inversion symmetry.
In many compounds it is the spiral spin configuration which breaks the
symmetry. However, as one can see from the calculation of KNB, this effect, to
the leading order, is only manifest in the spin part of the wave functions and
it bears the factor $\widehat{e}_{1}\times\widehat{e}_{2}$. The spatial part
of the wave functions, to the leading order, still possess space inversion
symmetry. Hence, the expectation values of electric polarization suffers
cancellation. There are other evidences that the lower the symmetry of the
system, the greater the electric polarization. In \cite{13} , it was found
that atom displacement enhanced electric polarization in E-type
antiferromagnetism. The results in ref. \cite{19} provide even stronger
support. Their calculation showed that the spin-orbit interaction is
indispensable to the existence of electric polarization while lattice
distortion enhances electric polarization by at least ten-fold. However, Bridges et. al. \cite{20} did not find evidence of lattice distortion within  experimental accuracy, which is $5\times 10^{-3}$\AA . It must be said that a displacement of O$^{2-}$ ions approximately $1\times 10^{-3}$\AA  is sufficient to give the electric polarization found in TbMnO$_{3}$. It will be even smaller for the displacement of Cu$^{2+}$ ions ($4\times 10^{-4}$\AA ) in the compounds with smaller polarization, like LiCu$_2$O$_2$ \cite{205}. At this stage, experimental data can neither support nor rule out lattice distortion contribution in TbMnO$_3$ and LiCu$_2$O$_2$. Therefore, it is likely that in the compounds with large electric
polarization such as TbMnO$_{3}$ and CuO \cite{21} there is an extra factor which reduce the symmetry. \textit{As a result, the spatial part of the
system also has lower symmetry to avoid the catastrophic cancellation}.

\textit{In this work we proposed that there should be a third long range order
besides the electric polarization and magnetic order.} It can be structure
distortion, charge order, or orbital order (OO). OO occurs in transition metal
compounds with crystal field, Jahn-Teller or GdFeO$_{3}$-type distortions
\cite{22}. For example in manganites, $\psi_{3x^{2}-r^{2}}$ and $\psi
_{3y^{2}-r^{2}}$ orbitals are favored on two inter-penetrating sublattices due
to crystal field \cite{23}. However, other effects such as superexchange
interaction and spin-orbit interaction can also change the orbital
configuration. In the spin-orbital model \cite{22} different combinations of
orbitals can give different exchange energies. Hence, there are various
competing mechanisms. As a result, the OO should not have such a simple
structure. We can provide another argument for the existence of OO. In
3d-transition elements the spin-orbit coupling strength is of the order 0.05eV
or 600K. It is much higher than the N\'{e}el temperature. This means that the
spin and orbital angular momenta are aligned in opposite direction long before
the magnetic order is established. If there is experimental evidence of say,
sinusoidal magnetic order, it is also an indication of OO with a wave vector
commensurate with that of spins.

Even if in a system without OO, structure distortion, such as cooperative
Jahn-Teller effect or GdFeO$_{3}$-type distortions can also play the role of
the third long order. The d-orbitals are anisotropic. The magnitude of oxygen
bond angle can influence hybridization. The third possible candidate is charge
order or its multipole expansion. For apparent reason it can cause an
additional periodic potential and hence, affects wave functions.

Here is a brief summary of this work. By introducing a third long range order
into the system, we show that the electric polarization can be enhanced by an
order of magnitude. A new form of electric polarization is derived. This long
range order can reduce the spatial symmetry of the spatial wave functions.
Therefore, the catastrophic cancellation is avoided. More specifically, long
range orders with wave vector $\vec{Q}$ affect the hybridization energy. The
state of crystal momentum $\vec{k}$ mixes with the state of $\vec{k}\pm\vec
{Q}$. In evaluation of electric polarization one has to sum over entire bands
up to the Fermi surface for multiferroic insulators. The cancellation will
\textit{not} be complete if the wave function is a mixture.

To see the origin of the catastrophic cancellation we start with an
U(1)$\times$SU(2) gauge-invariant action which will give the equation of
motion as the Pauli equation:
\begin{align}
S  &  =\int d^{4}x[i\hslash\psi^{\ast}D_{0}\psi-\frac{\hslash^{2}}{2mc}%
\sum_{k=1}^{3}(D_{k}\psi)^{\ast}(D_{k}\psi)\nonumber\\
&  -\frac{1}{16\pi c}\sum_{\mu,\nu=0}^{3}F_{\mu\nu}F^{\mu\nu}] \label{eq2}%
\end{align}
with $x^{0}=ct$ and the covariant derivatives $D_{0}=\partial/\partial
x^{0}-(ie/\hslash c)A_{0}+(ie/2mc^{2})\sum_{a}\rho_{0a}\sigma^{a}$ and
$D_{k}=\partial/\partial x^{k}+(ie/\hslash c)A_{k}+(ie/4mc^{2})\sum\rho
_{ka}\sigma^{a}$ are introduced in order to show clearly the charge and spin
currents. A term of EM field is also present. Here, we use the notations of
Fr\"{o}hlich and Studer \cite{24} with a little simplification. The SU(2)
fields $\rho_{0a}=B^{a}$ and $\rho_{ka}=\sum_{b}\varepsilon_{kab}E^{b}$ are
related the components of magnetic and electric field. However, it should be
noted that the action is correct only to the order $O(m^{-2})$. Above action
illustrates the origin of AC effect in which the electric field is coupled to
moving spins or magnetic moments. By taking variation of $\rho_{ka}$,
multiplying by $\varepsilon_{kab}$ and summing over $a$, and assuming the
system is charge neutral, we obtain
\begin{align}
\frac{E_{b}}{4\pi c}  &  =\frac{-\hslash e}{8m^{2}c^{3}}{\sum\nolimits_{a,k=1}%
^{3}}\{[(\frac{\hslash}{i}\frac{\partial}{\partial x^{k}}+\frac{e}{c}%
A_{k})\psi]^{\ast}\varepsilon_{kab}\sigma^{a}\psi\nonumber\\
&  +(\varepsilon_{kab}\sigma^{a}\psi)^{\ast}(\frac{\hslash}{i}\frac{\partial
}{\partial x^{k}}+\frac{e}{c}A_{k})\psi\} \label{eq3}%
\end{align}
where a term of the order $m^{-3}$ is discarded. The r.h.s. is clearly the
spin current. It is the gist of the spin current model of multiferroics. From
the equation $\vec{D}=\vec{E}+4\pi\vec{P}$ one finds that when there is no
free charge, the electric polarization $\vec{P}=\vec{E}/4\pi$ can be generated
by spin current. Nevertheless, we can see the reason why the electric
polarization given by the spin current model usually is small in magnitude. In
insulators such as multiferroics, there is no movement of free charge carriers
and thus no spin current on the r.h.s. of Eq.(\ref{eq3}). KNB had to invoke
superexchange interaction. This is exactly the reason why there is
catastrophic cancellation because only by going to the higher orders can one
find spin current. The cancellation can be avoided if there is a long range
order in the system which can give the wave function modulation, produce
significant derivative of $\psi$ and hence, give rise to large the electric polarization.

Our model is quite simple. A crystal of transition element atoms with valence
d-orbitals and oxygen atoms with p-orbitals are considered. The transition
metal atoms have spiral spin configuration and OO so that the j-th transition
metal atom has the combined spin and orbital state
\begin{align}
\Psi_{j}  &  =%
\begin{pmatrix}
e^{-i\phi/2}\cos(\vec{q}\cdot\vec{R}_{j}/2)\\
e^{i\phi/2}\sin(\vec{q}\cdot\vec{R}_{j}/2)
\end{pmatrix}
\nonumber\\
&  \otimes%
\begin{pmatrix}
\lbrack A+B\cos(\vec{Q}\cdot\vec{R}_{j})]\psi_{d1}(\vec{r}-\vec{R}_{j})\\
B\sin(\vec{Q}\cdot\vec{R}_{j})\psi_{d2}(\vec{r}-\vec{R}_{j})
\end{pmatrix}
\label{eq4}%
\end{align}
where $\psi_{d1(2)}(\vec{r}-\vec{R}_{j})$ is the first and second d-orbital of
the transition element atom at site $\vec{R}_{j},$ $\vec{q}$ and $\vec{Q}$
stand for the wave vectors of spin order and orbital order respectively and
$\phi$ is the angle between x-axis and the spin. We assume that there are two
active orbitals like those of e$_{g}$ orbitals and have reserved the freedom
of OO with non-orthogonal orbitals, like $\left\vert 3x^{2}-r^{2}\right\rangle
$ and $\left\vert 3y^{2}-r^{2}\right\rangle $. $B$ is approximately a
constant. If on the other hand, the third long-range order is lattice
distortion and there is no OO, then $A=1$\ and $B=0,$ (or one can set $Q=0.$).
The effect of lattice distortion will appear in the Hamiltonian in next
paragraph in the form of modulated hybridization energy.

The hybridization energy is affected by the orbital states and lattice
distortion due to the anisotropies of the orbitals. It is also influenced by
spins. The spins of the hopping electrons are aligned with localized spins.
Hence, the Hamiltonian has the form%
\begin{align}
H  &  =\sum\varepsilon_{p}c_{pi,l\sigma}^{\dagger}c_{pi,l\sigma}%
+\sum\varepsilon_{d}c_{dj}^{\dagger}c_{dj}\nonumber\\
&  +\sum_{n.n.}[AV_{l}+BV\cos(\vec{Q}\cdot\vec{R}_{j}-\alpha_{l})]\cos(\vec
{Q}^{\prime}\cdot\vec{R}_{j})\nonumber\\
&  [e^{-i\phi/2}\cos\theta_{j}c_{dj}^{\dagger}c_{pi,l\uparrow}+e^{i\phi/2}%
\sin\theta_{j}c_{dj}^{\dagger}c_{pi,l\downarrow}]+H.c.\nonumber\\
&  +\lambda\sum\vec{l}_{j}\cdot\vec{s}_{j} \label{eq5}%
\end{align}
where $l$\ is the index of the oxygen atoms (position $\vec{R}_{j}+\vec{r}%
_{l}$) in the basis, and $\theta_{j}=\vec{q}\cdot\vec{R}_{j}/2.$ $V_{l}$\ and
$V_{l}^{\prime}$ are the hybridization energies of the p-orbital of the $l$-th
oxygen atoms with the first and second d-orbital of the transition metal
atoms, $V=\sqrt{V_{l}^{2}+V_{l}^{\prime2}}$ and $\cos\alpha_{l}=V_{l}/V.$ The
last term is the spin-orbit interaction. The factor $AV_{l}+BV\cos(\vec
{Q}\cdot\vec{R}_{j}-\alpha_{l})$ in hybridization comes entirely from OO.\ The
modulation $BV\cos(\vec{Q}\cdot\vec{R}_{j}-\alpha_{l})$ contains contribution
from the first and second orbitals with the factors $\cos\alpha_{l}$\ and
$\sin\alpha_{l}$\ respectively.\ If there is no OO, then $A=1$, $B=0,$
$\alpha_{l}=0$. The factor $\cos(\vec{Q}^{\prime}\cdot\vec{R}_{j})$ comes from
the displacements of atoms. For example, if the oxygen atom\ between two
transition metal atoms\ move laterally such that the bond angle of TM-O-TM is
reduced, then the hybridization will be reduced for e$_{g}$ electrons. Hence
the hybridization energy acquires the modulation of lattice distortion (wave
vector $\vec{Q}^{\prime}$). If both OO and lattice distortion are present,
then our case is viable only if they are commensurate to each other. In that
case, two effects are superimposed on each other and the hybridization energy
will has the form above. In the following calculation we will consider the
case with OO only. The extension to additional lattice distortion effect is
not difficult.

The Hamiltonian without spin-orbit interaction can be diagonalized by making
the substitution: $\vec{R}_{i,l}=\vec{R}_{i}+\vec{r}_{l}$, $p_{il,\uparrow
}=e^{-i\phi/2}\cos(\vec{q}\cdot\vec{R}_{i,l}/2)c_{pi,l\uparrow}+e^{i\phi
/2}\sin(\vec{q}\cdot\vec{R}_{i,l}/2)c_{pi,l\downarrow}$ and $p_{il,\downarrow
}=e^{i\phi/2}\cos(\vec{q}\cdot\vec{R}_{i,l}/2)c_{pi,l\downarrow}-e^{-i\phi
/2}\sin(\vec{q}\cdot\vec{R}_{i,l}/2)c_{pi,l\uparrow}$, and taking Fourier
transform. For our purpose, it suffices to assume that $\varepsilon
_{d}-\varepsilon_{p}>|V|$ and analyze the perturbed wave function. The
resulting wave function can be expressed with the following tight-binding wave
functions: $\psi_{d,\vec{k}}(\vec{r})={\displaystyle\sum\nolimits_{j}}%
e^{i\vec{k}\cdot\vec{R}_{j}}\Psi_{j}$ and $\psi_{p,l,\vec{k}}(\vec
{r})={\displaystyle\sum_{j}}e^{i\vec{k}\cdot\vec{R}_{j}}\psi_{p}(\vec{r}%
-\vec{R}_{j}-\vec{r}_{l})(e^{-i\phi/2}\cos\theta_{j},e^{i\phi/2}\sin\theta
_{j})^{T}$and it is
\begin{align}
\psi_{\vec{k}}(\vec{r})  &  =C\Big\{\psi_{d,\vec{k}}(\vec{r})+{ \displaystyle\sum
\nolimits_{l}}\frac{V_{l}}{\varepsilon_{d}-\varepsilon_{p}}\Bigl[A\psi_{p,l,\vec
{k}}(\vec{r})\nonumber\\
&  +{\displaystyle\frac{B}{2}\sum\nolimits_{\zeta}}e^{-i\zeta\alpha_{l}}%
\psi_{p,l,\vec{k}+\zeta\vec{Q}}(\vec{r})\Bigr] \Big\} \label{eq6}%
\end{align}
where $\psi_{p}(\vec{r}-\vec{R}_{j}-\vec{r}_{l})$ is the p-orbitals of the
oxygen atoms at site $\vec{R}_{j}+\vec{r}_{l}$, $\zeta=\pm1$, and $C$ is a
normalization constant.

We apply the spin-orbit interaction to the wave functions in Eq.(\ref{eq6}) as
a perturbation and gain a term $-(\lambda/\Delta)\vec{l}\cdot\vec{s}\Psi_{j}$
where $\Delta=J_{H}-\Delta_{cf}$ with $J_{H}$ being the Hund's coupling energy
and $\Delta_{cf}$ the crystal field splitting between e$_{g}$ and t$_{2g}$
orbitals. (The perturbed states have minority spin and t$_{2g}$ orbitals.) Now
we are in position to calculate the electric polarization with Eq.(\ref{eq6}):
\begin{align}
\vec{P}  &  =\frac{e|C^{2}|\lambda}{(\varepsilon_{d}-\varepsilon_{p})\Delta
}{\displaystyle\sum\nolimits_{\vec{k},l}}V_{l}\Bigl[ a\langle\psi_{d,\vec{k}}%
|(\vec{l}\cdot\vec{s})\vec{r}|\psi_{p,l,\vec{k}}\rangle\nonumber\\
&  +{\displaystyle\sum\nolimits_{\zeta}}\frac{be^{-i\alpha_{l}\zeta}}%
{2}\langle\psi_{d,\vec{k}}|(\vec{l}\cdot\vec{s})\vec{r}|\psi_{p,l,\vec
{k}+\zeta\vec{Q}}\rangle \Bigr] +c.c.. \label{eq7}%
\end{align}
The unperturbed wave functions do not contribute to electric polarization
because of cancellation. It is due to crystal structure inversion symmetry
despite of spiral spin configuration. The first term vanishes after summing
over $\vec{k}$. It is again due to the inversion symmetry of the spatial part
of the wave function. The net electric polarization comes from the second term
in Eq.(\ref{eq7}). The reason the catastrophic cancellation is avoided is that
with the third long range order, the energy bands open gaps at $2\vec{k}%
\cdot(\vec{Q}+\vec{G})=(\vec{Q}+\vec{G})^{2}$ where $\vec{G}$ is a reciprocal
lattice vector. The gap is comparable to the hybridization energy if OO or
lattice distortion is significant. Different mini-bands not only have
different energies but also have different linear combinations of wave
functions. In other words, the symmetry of the entire band is lower.
Introducing phonons into the system \cite{19} has the same effect. The
importance of $\vec{Q}$ is manifest. Calculation of the contribution of the
highest mini-band gives
\begin{align}
\vec{P}  &  \simeq\widehat{e}_{z}\frac{4eI^{\prime}AB\lambda V\cos\alpha}%
{\pi\Delta(\varepsilon_{d}-\varepsilon_{p})}\nonumber\\
&  [\cos(Q_{x}a_{0})\sin(Q_{x}a_{0}/2)\sin(q_{x}a_{0}/2)\cos\phi\nonumber\\
&  +\cos(Q_{x}a_{0})\sin(Q_{y}a_{0}/2)\sin(q_{y}a_{0}/2)\sin\phi]. \label{eq8}%
\end{align}
where $a_{0}$ is the lattice constant and $I^{\prime}={\int}\psi_{zx}^{\ast
}(\vec{r}-\vec{R}_{j})z\psi_{x}(\vec{r}-\vec{R}_{j}-\vec{r}_{l})d^{3}\vec{r}$.
We define $\cos\alpha=|\cos\alpha_{l}|,$\ note that for $x^{2}-y^{2}$\ orbital
$V_{x}=-V_{y}$\ and for $3z^{2}-r^{2}$\ orbital, $V_{x}=V_{y}$, so that
$\cos\alpha_{x}=\pm\cos\alpha_{y}.$\ In the continuum limit where
$a_{0}\rightarrow0$, we found
\begin{equation}
\vec{P}\simeq-\frac{2eI^{\prime}V\lambda}{\pi(\varepsilon_{d}-\varepsilon
_{p})\Delta}\widehat{Q}\times(\widehat{s}_{j}\times\widehat{s}_{j+1}).
\label{eq9}%
\end{equation}
Here, $\widehat{s}$ is the unit vector of the spin. In a crystal with a
long-range order other than spins and electric polarization, the wave vector
of the long-range order $Q$, emerges in the expression of electric
polarization in place of the bond direction in Eq.(\ref{eq1}), as shown in
Fig. 1.

\begin{figure}[t]
\begin{center}
\includegraphics[scale=0.5]{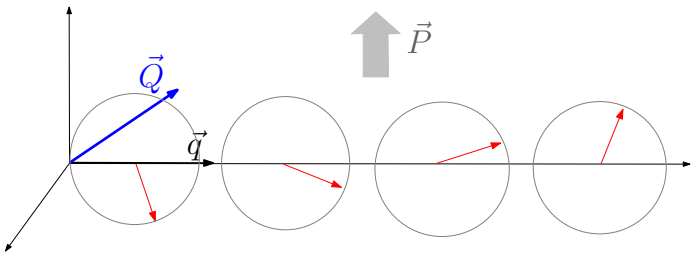}
\end{center}
\caption{(color online) Schematic graph showing the directions of spins (red
arrows), electric polarization (grey arrow) $\vec{q}$ (black arrow) and
$\vec{Q}$ (blue arrow). $\vec{q}$ and $\vec{Q}$ can be parallel to each other.
See text.}%
\end{figure}

Since we are concerned with the magnitude of the electric polarization, we now
make a practical estimation: $|V_{l}|\approx0.5eV,$ $\varepsilon
_{d}-\varepsilon_{p}\approx\Delta\approx1eV$, $\lambda\approx0.05eV$ and
$AB\thickapprox0.5.$ Note that above calculation is purely of hybridization.
It does not involve double occupancy in the transition metal atoms and hence,
the on-site Coulomb repulsion does not have any effect. For this reason
$\varepsilon_{d}-\varepsilon_{p}$ and $\Delta$ are smaller than those
calculated with LDA+U density functional theory. The magnitude of electric
polarization is $|\vec{P}|\sim3000-5000(I^{\prime}/ \operatorname{\text{\AA}}%
$)$\mu C/m^{2}$. Since the order of magnitude of $I^{\prime}$ is a fraction of
$\operatorname{\text{\AA}}$. The result of our calculation is compatible with
the experimental data of say, TbMnO$_{3}$.

On reflection of Eq.(\ref{eq3}), one sees clearly the origin of the electric
polarization in Eq.(\ref{eq9}). The $\psi$\ in Eq.(\ref{eq3}) is just that in
Eq.(\ref{eq6}). The scattering of electrons with the third long range order,
which can be seen from our calculations in Eq.(\ref{eq4}-\ref{eq6}), mixes
different plane waves and hence, enable $\psi$\ to sustain current. With the
magnetic moments provided by spiral spins in leading order, the spin current
on the r.h.s. of Eq.(\ref{eq3}) gives rise to electric field on the l.h.s.
which in turn, causes electric polarization in the form $-\overrightarrow
{E}\cdot\overrightarrow{P}$. With linear response formalism
\begin{equation}
\overrightarrow{P}=-e^{2}\sum_{n}\frac{\langle0|\overrightarrow{E}%
\cdot\overrightarrow{r}|n\rangle\langle n|\overrightarrow{r}|0\rangle}%
{E_{0}-E_{n}} \label{eq10}%
\end{equation}
one finds that $\overrightarrow{P}$\ in Eq.(\ref{eq9}) and Eq.(\ref{eq10}%
)\ has the same form. We have the following correspondence: $E_{n}-E_{0}%
\sim\Delta$,\ $\overrightarrow{E}\cdot\overrightarrow{r}$ gives spin-orbit
interaction with $\overrightarrow{E}$\ in Eq.(\ref{eq3}) and the spin operator
in Eq.(\ref{eq3}) gives the now well-quoted factor $\widehat{s}_{j}%
\times\widehat{s}_{j+1}$. Since spins and orbitals are closely related, wave
vectors of magnetic orders and orbital order very likely have the relation
$\vec{Q}=r\vec{q}$\ where $r$ is a rational number. Our calculation showed
that only incommensurate OO can suppress catastrophic cancellation. If not, OO
can be accounted for with a enlarged unit cell, and the electric polarization
still suffers catastrophic cancellation. This can explain why magnetoelectric
phenomenon usually occurs with incommensurate magnetic orders.

Our result has several implications. For example, it was found that
\cite{25,18} there is modulated electric polarization (antiferroelectricity)
in multiferroics. Its absolute magnitude is much greater than that of the net
polarization. The existence of OO makes things more complex. If the modulated
polarization and OO are commensurate to each other, then an enhancement can
occur. A related phenomenon is electromagnons \cite{26} in multiferroic
material \cite{27}. It was suggested \cite{28} that the dominant term is of the form
$\vec{P}=\sum\vec{\Pi}_{ij}\vec{S}_{i}\cdot\vec{S}_{j}$. According to Moriya
\cite{26}, $|\vec{\Pi}_{ij}|\sim I^{\prime}J/\Delta$\ where the superexchange
interaction $J$\ ($V$ in Moriya's notation) is of the order $(V^2/\Delta)^2/U%
$\ with $U$\ being the on-site Coulomb repulsion. Comparing to this\ $\vec
{P},$\ our polarization in Eq.(\ref{eq9}) is more favorable or the spin-orbit
coupling (0.05eV) is relatively stronger compared to exchange interaction. The
additional long range order can give rise to extra magnon excitations. These are of purely electronic origin. The wave vector of the magnons can also be different. It is equal to $\vec{Q}$ for one-magnon processes and $\vec{k}$ and $\vec{k}\pm \vec{Q}$ for bi-magnon processes. Its effect will be manifest in optical spectra, susceptibility and
electric polarizability measurement. This will be related to the vector spin
chirality \cite{29} $\widehat{s}_{j}\times\widehat{s}_{j+1}$. The study on
this subject is under way.

In conclusion we have proposed that with third long range order in the system,
the catastrophic cancellation of electric polarization can be avoided. The
resulting magnitude is comparable to experimental data. The long range order
can be either OO or cooperative Jahn-Teller distortion, but it has to be
incommensurate. Lastly but perhaps most importantly, the cause of electric
polarization and multiferroics is firmly established and it is spin current
coming from the spin-orbit interaction.

The author is in debt to discussion with G. Y. Guo and N. Nagaosa. He also
benefits from the activities of "quantum novel phenomena in condensed matter"
focus groups of NCTS, Taiwan. This work is supported in part by the National
Science Council under the contract NSC 98-2112-M-002-011-MY3.

\end{document}